\documentclass[doublecol]{epl2} 

\title{Thinning or thickening? Multiple rheological regimes 
in dense suspensions of soft particles}

\shorttitle{Thinning or thickening?}

\author{Takeshi Kawasaki \and Atsushi Ikeda \and 
Ludovic Berthier}

\institute{Laboratoire Charles Coulomb, UMR 5221, CNRS and Universit\'e
Montpellier 2, Montpellier, France}


\abstract{The shear rheology of dense colloidal and granular 
suspensions is strongly nonlinear, as these materials exhibit 
shear-thinning and shear-thickening, depending on 
multiple physical parameters. We numerically study the 
rheology of a simple model of soft repulsive
particles at large densities, and show 
that nonlinear flow curves reminiscent of experiments on real 
suspensions can be obtained. By using dimensional analysis 
and basic elements of kinetic theory, we rationalize these 
multiple rheological regimes and disentangle the relative impact 
of thermal fluctuations, glass and jamming transitions, 
inertia and particle softness
on the flow curves. We characterize more specifically the shear-thickening 
regime and show that both particle softness and the emergence of a 
yield stress at the jamming transition compete with 
the inertial effects responsible for the observed thickening behaviour. This 
allows us to construct a dynamic state diagram, which can be used 
to analyze experiments.} 

\usepackage{epsfig,amssymb,amsfonts,amsmath}
\usepackage{graphicx}
\newcommand{\be}{\begin{equation}}
\newcommand{\ee}{\end{equation}}
\newcommand{\gdot}[0]{\dot{\gamma}}

\renewcommand{\phi}{\varphi}
\newcommand{\odif}[2]{\frac{{\rm d} #1}{{\rm d} #2}}
\newcommand{\pdif}[2]{\frac{\partial #1}{\partial #2}}
\newcommand{\ba}{\begin{align}}
\newcommand{\ea}{\end{align}}

\usepackage{color}

\begin{document}

\maketitle

\section{Introduction} 

Understanding the shear rheology of dense 
colloidal and granular suspensions remains a central challenge 
at the crossroad between nonequilibrium statistical mechanics
and soft matter physics, with a clear technological 
relevance~\cite{larson,coussot,wagner}. 
Simple liquids display simple rheological properties
characterized by linear Newtonian behavior~\cite{hansen}. 
In  a simple shear flow,
for instance, the rate of deformation, $\gdot$, is proportional to the applied 
shear stress, $\sigma$, such that the viscosity $\eta = \sigma / \gdot$
uniquely characterizes the rheological response.
 
However, in dense particle suspensions such as emulsions, 
colloidal assemblies,
or granular materials, the viscosity is usually not a single number, 
but a nonlinear function of the applied flow rate. To characterise
these materials, an entire {flow curve} $\eta = \eta(\gdot)$ 
is thus needed~\cite{larson,coussot,wagner}. Because the 
applied deformation now determines the response of the system, understanding
nonlinear flow curves obviously requires a more detailed 
analysis, which must deal with both nonlinear and nonequilibrium 
effects. When the viscosity varies with the applied shear rate, 
the system can either flow more easily as $\gdot$ 
increases (shear-thinning), or offer increasing resistance to flow
(shear-thickening). 
We are familiar with both these effects, as most complex 
fluids used for cosmetics or in food products display these 
amusing nonlinearities, which can be
technologically both useful or annoying~\cite{wagner,wagnerpt}.  

\begin{figure}
\psfig{file=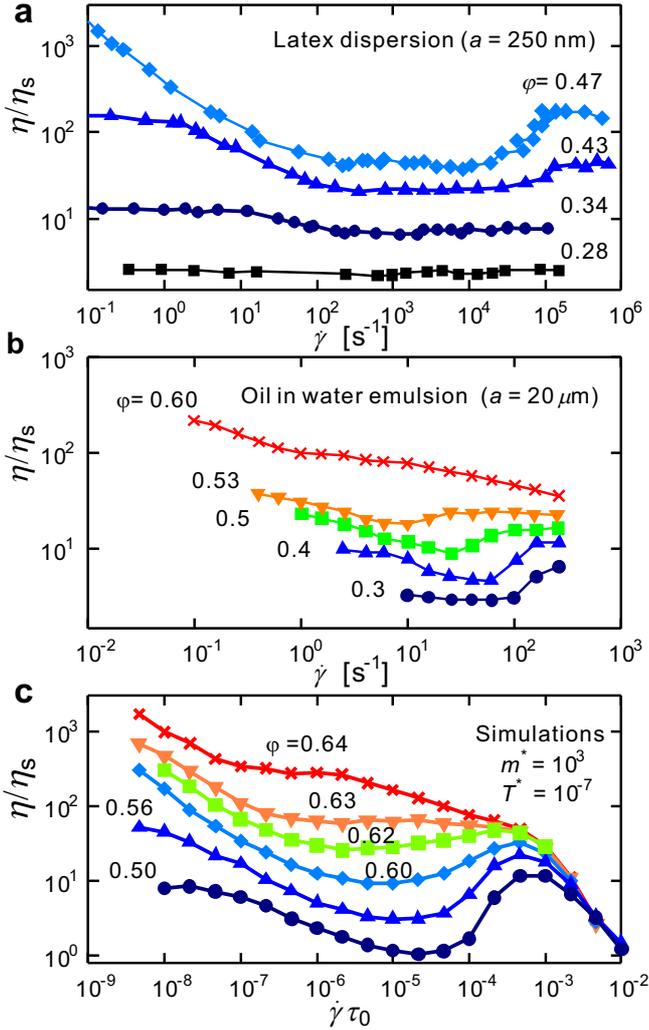,width=8.5cm,clip}
\caption{\label{fig1} 
Flow curves $\eta = \eta(\gdot)$ for various volume fractions
$\phi$ obtained in two experimental systems (a,b) and
measured in our numerical simulations (c).
(a) Latex dispersion  with diameter 250 nm~\cite{laun,wagnerpt}. 
(b) Oil-in-water emulsion with diameter 20 $\mu$m~\cite{otsubo}.
(c) Numerical simulations of harmonic spheres temperature and inertia
included using adimensional parameters $T^\star = 10^{-7}$ and $m^\star = 10^3$. 
All three systems reveal a coexistence of multiple Newtonian, shear-thinning,
and shear-thickening regimes, in a succession that strongly 
depends on volume fraction.}
\end{figure}

In practice, most of experimental flow curves measured 
even in model suspensions display a complex mixture of both these 
nonlinear effects~\cite{coussot,wagner,wagnerpt}. As two typical 
examples, we show flow curves measured 
in a colloidal dispersion of latex particles~\cite{laun,wagnerpt} (diameter 
$a=250~{\rm nm}$, Fig.~\ref{fig1}a), 
and in an oil-in-water emulsion~\cite{otsubo} (diameter $a=20~\mu$m,
Fig.~\ref{fig1}b). For a given volume fraction $\phi$, the flow curves 
may display an initial Newtonian regime at low enough $\gdot$
and $\phi$, or a strong shear-thinning regime when $\phi$ is larger.
This thinning regime is followed, for intermediate $\phi$ and larger $\gdot$,
by a Newtonian plateau regime. At larger $\gdot$, shear-thickening 
sets in, and the magnitude of the viscosity increase clearly  
depends on the density regime. 
In some cases, shear-thickening is interrupted and 
flow curves display a viscosity maximum. Finally
shear-thickening is not observed when density is too large, see
for instance the large density data in Fig.~\ref{fig1}b.  
Books and reviews of course offer an even broader range of possible 
behaviours~\cite{larson,wagner,coussot,wagnerpt,Barnes_review,jbreview}, 
but the data in Figs.~\ref{fig1}a,b are 
representative of the typical behaviour of dense suspensions.  

The primary purpose of this work is to show that a simple model of 
soft repulsive particles can exhibit a similarly complex rheology, 
despite the fact that {\it it does not incorporate} 
several of the physical ingredients usually put forward to account for
these nonlinearities. We argue that understanding first such a simple model 
is useful before turning to more complicated physical explanations
which can only increase the complexity of the results.  
In our model of spherical particles, we use a canonical pairwise, 
purely repulsive particle interaction. Dynamics is controlled 
by a traditional Langevin dynamics, which we simulate at 
either finite or zero temperature. 
The system is sheared at constant shear rate $\gdot$ in a simple 
shear geometry, it flows homogeneously (no shear-banding) in the absence of 
any kind of density or stress gradients. We do not introduce more 
complex ingredients 
such as hydrodynamic interactions, attractive forces, lubrication forces, 
or frictional contacts between the particles. Our model, we believe, 
represents an ideal starting point to understand the basic physics
of suspensions, to which more ingredients could then be added, if needed. 

In Fig.~\ref{fig1}c, we present a set of flow curves 
obtained from our simulations, for a specific set of 
parameters, which we shall discuss shortly. 
For moderate volume fractions, we obtain a complex
succession of Newtonian, shear-thinning, Newtonian, shear-thickening, 
shear-thinning regimes as the flow rate is varied. The first Newtonian 
regime disappears as $\phi$ is increased, and at even larger 
$\phi$ the intermediate Newtonian and thickening regimes 
also disappear, such that the flow curve becomes purely shear-thinning
at large density. The existence of these 
multiple rheological regimes and their evolution with volume 
fraction is in excellent qualitative agreement with the experimental 
flow curves presented in Figs.~\ref{fig1}a,b. Given the extreme simplicity of
our model, we believe that such an agreement is an important
achievement of our work.

In the following, we first summarize previous knowledge on 
nonlinear flow curves in dense suspensions. We then discuss our 
numerical results using dimensional analysis and basic elements of 
kinetic theory, and provide a simple understanding of all the regimes observed 
numerically. 

\section{Thinning and thickening}

Shear-thinning is the most commonly observed 
nonlinearity in experimental flow curves~\cite{larson}. This 
effect typically results from 
the competition between the complex structure of the suspension under study 
that is responsible for the large viscosity of the system at rest, 
and the flow rate which tends to disrupt this 
static organization. For dense suspensions, large viscosities 
emerge when either the glass transition (for thermal systems~\cite{rmp}), 
or the jamming transition (for athermal ones~\cite{reviewmodes}) is approached. 
In both cases, shear-thinning occurs when the shear rate 
competes with the relevant timescale associated to structural relaxation. 

By contrast, shear-thickening is typically less common in 
experiments~\cite{Barnes_review}.
Additionally, its detailed characterization is more difficult, because
shear-thickening is often mixed with issues such as flow heterogeneities, shear
bands, density gradients, or particle migration~\cite{bonn,fall}. 
However, shear-thickening has been documented for a large number of 
dense suspensions of various types, for soft and hard 
particles, for small colloids and large granular 
particles~\cite{wagner,wagnerpt,Barnes_review}.
This variety of systems has led to a similar variety of 
theoretical arguments and models to account for 
shear-thickening, but a broad consensus has yet to emerge. 
In some systems, a strong increase of the viscosity can even become
a nearly discontinuous jump, which we do not observe in our 
simple system. Another regime that we do not explore is 
suspensions at moderate densities, where a shear-thickening 
of modest amplitude likely results from 
hydrodynamic interactions promoting cluster formation 
under shear~\cite{brady,wagnerpt}. 

Recent experimental work conducted in dense suspensions 
shows that a shear-thickening of large amplitude is 
observed in the vicinity of the jamming transition, suggesting 
that particle crowding promotes 
shear-thickening~\cite{jbreview,brownprl,brownjr,fall,hayakawa,claudin,seto,seto2,fernandez,xu,claus}. 
However, the emergence of a yield stress above the jamming transition 
was argued to suppress shear-thickening~\cite{brownprl,jbreview},
suggesting that some crowding is needed, but not {\it too much} crowding.
These observations suggest that neglecting hydrodynamic
interactions to focus instead on the specific competition 
between steric constraints and the shear flow is an interesting path. 

Recent work~\cite{fall,fernandez,claudin} explored the idea 
that a strong continuous increase of the viscosity under shear 
results from a change of dissipation mechanism from 
viscous to inertial damping, in the spirit of 
Bagnold~\cite{bagnold}.  We extend these 
ideas to include also the effect of thermal fluctuations and particle 
softness, which are experimentally relevant.
In recent work, we analysed in detail 
the complex interplay of thermal fluctuations and steric constraints
in the overdamped limit~\cite{ikeda,ikedasoft}. Here we add inertia,
which introduces another timescale and adds a new level 
of complexity to flow curves that are already nontrivial.  
 
For completeness, we mention recent studies 
invoking frictional forces to explain discontinuous 
shear-thickening~\cite{hayakawa,brownjr,seto,seto2,fernandez,claus}. 
Numerical studies seem in good qualitative agreement with experiments, although
quantitative understanding and detailed comparison to 
experimental data are unavailable.

\section{Model and numerical simulations} 
  
We study a system composed of $N$ particles with equal mass $m$, 
interacting through a purely repulsive truncated 
harmonic potential~\cite{durian}, $V(r) = \frac{\epsilon}{2} (1 - r/a)^2 
\Theta (a-r)$, 
where $a$ is particle diameter, $\epsilon$ an energy scale, 
and $r$ the interparticle distance; $\Theta(x)$ is the Heaviside
function. In practice, to avoid crystallization which might occur 
at large density, we use a $50:50$ binary mixture of spheres with 
diameter ratio 1.4. The resulting volume fraction is 
$\phi = \frac{\pi}{12L^3} N (a^3 + (1.4 a)^3)$, where $L$ 
is the linear size of the simulation box. 
We use $N=10^3$ particles in a cubic box in three spatial dimensions, 
and inforce Lees-Edwards periodic boundary conditions~\cite{allen-tildesley}.
 
These soft particles evolve with Langevin dynamics, 
\be
m\odif{\vec{v}_i}{t} + \xi (\vec{v}_i - \gdot y_i\vec{e}_x) + \sum_{j \neq i} 
\pdif{V(|\vec{r}_i - \vec{r}_j|)}{\vec{r}_i} + \vec{f}_i = 0, \label{eom} 
\ee
where $\vec{r}_i$ and $\vec{v}_i$ respectively 
represent the position and velocity of particle $i$. 
The first term is the particle acceleration, 
and the second represents the viscous damping,  
controlled by the damping constant $\xi$. The system 
is sheared in the $xy$ plane, and advection occurs along 
the $x$ axis. The shear rate is $\dot{\gamma}$, 
$y_i$ is the $y$ coordinate of particle $i$, 
and $\vec{e}_x$ is the unit vector along the $x$ axis.
The third term in Eq.~(\ref{eom}) incorporates 
pairwise harmonic repulsion between the particles, 
and the final term  
is the Brownian random force acting on particle $i$, 
which we draw from a Gaussian distribution with zero mean
and variance obeying the fluctuation-dissipation relation, 
$\langle \vec{f}_i(t)\vec{f}_j(t')^{T}\rangle=2k_BT\xi\delta_{ij}{\bf 1} 
\delta(t-t')$,
where $k_{\rm B}$ is the Boltzmann constant and $T$ the temperature.

We impose the shear rate $\gdot$ and measure the shear 
stress $\sigma$ as the  
$xy$ component of stress tensor using the 
Irving-Kirkwood formula~\cite{allen-tildesley}. 
We deduce the shear viscosity, $\eta = \sigma/\gdot$. 

\section{Relevant timescales and units}

While very simple, Eq.~(\ref{eom}) contains a number of 
distinct ingredients which typically 
control the physics of dense suspensions: 
inertial forces, viscous damping, particle interactions,  
and thermal noise. Their competition is more easily understood 
by introducing characteristic timescales to each term.
We can define three independent timescales. 
The damping time of particle velocity is given 
by $\tau_v = m/\xi$. Energy is dissipated through viscous 
damping over a timescale $\tau_0 = \xi a^2 /\epsilon$, 
while thermal fluctuations occur over a Brownian timescale 
$\tau_T = \xi a^2 /(k_BT)$.

We can construct two dimensionless parameters 
out of these three timescales, which we will use 
to quantify the relative effects of inertia, thermal 
fluctuations and viscous damping.
Comparing $\tau_v$ to $\tau_0$, we can create 
a dimensionless particle mass, 
$m^\star = \tau_v / \tau_0 = m \epsilon / (\xi a)^2$.
In the same spirit, comparing $\tau_T$ to $\tau_0$
allows the definition of a dimensionless temperature, 
$T^\star = \tau_T / \tau_0 = k_BT / \epsilon$.
We use both $T^\star$ and $m^\star$ to specify the values
of the control parameters employed in a given study, which amounts 
to choosing $\tau_0$ as our microscopic time unit.  
Therefore a set of simulations is fully specified by the values 
of $(m^\star, T^\star)$, for which we can then vary 
both the packing fraction $\phi$ and the flow rate $\gdot$. 

We use the `solvent' viscosity as a unit of viscosity, 
which is defined through the Stokes law, 
$\eta_s = \xi/ (3 \pi a)$. Accordingly, the natural
stress scale is $\sigma_s = \epsilon / a^3$. 

\section{Multiple regimes in the flow curves} 
  
Having defined the relevant timescales, we are now in a position 
to properly identify and interpret 
the multiple rheological regimes observed 
in the simulation results presented in Fig.~\ref{fig1}c,
which were obtained by fixing $m^\star = 10^3$ and $T^\star = 10^{-7}$.
As should now be obvious, these values imply a clear separation
of the relevant timescales, namely $\tau_T \gg \tau_v \gg \tau_0$.
As a result, the imposed shear rate successively competes 
with all three timescales, which directly impacts the measured flow
curves at various densities shown in Fig.~\ref{fig1}c, as we now explain. 

At $\varphi \leq 0.56$ a Newtonian behavior is observed in the limit
$\gdot \to 0$, which gives way to shear-thinning behaviour 
as $\gdot$ is increased. Both these behaviours
occur in the regime $\gdot \tau_T \lesssim 1$, which implies that
thermal fluctuations control this regime. ($\gdot \tau_T$ is 
the `bare' P\'eclet number). The corresponding 
Newtonian viscosity, $\eta_T$, is therefore directly related to 
the equilibrium structural relaxation timescale, $\tau_\alpha(T,\phi)$, 
of the thermalized suspensions of harmonic spheres~\cite{tom}, which 
can be calculated through equilibrium
linear response theory.
Instead, shear-thinning is characteristic of the 
nonlinear rheology of viscous glassy fluids~\cite{rmp}. 
It occurs when structural relaxation is provoked by the 
imposed shear flow (i.e. when the `dressed' P\'eclet number 
$\gdot \tau_\alpha$ is not small).   
 
At higher shear rate, $\gdot \tau_T \gtrsim 1$, thermal 
fluctuations cannot affect the physics which thus becomes 
equivalent to zero-temperature (or `athermal') rheology~\cite{ikeda}. 
In this regime, we first observe a Newtonian behavior,
with an apparent viscosity $\eta_0$ which has a different value than 
in the first Newtonian regime, $\eta_0 \neq \eta_T$. 
The difference between these thermal and athermal Newtonian regimes 
for the overdamped case with $m^\star =0$ is discussed in Ref.~\cite{ikeda}. 
Comparison with these earlier results reveals 
perfect agreement, which is expected 
as long as inertial effects do not affect the physics, 
i.e. when $\gdot  \tau_v$ is small. 

Increasing further the shear rate, 
deviations from the overdamped limit start to 
appear when $\gdot \tau_v$ becomes large~\cite{fall,claudin}. In this 
regime, we observe a succession of continuous shear-thickening
followed by a shear-thinning regime. As a result, 
the viscosity exhibits a maximum at a well-defined $\gdot$ value. 
The shear-thickening and the viscosity 
maximum are not present in the overdamped simulations~\cite{ikeda}, 
and are analysed in more detail below. 
 
Having properly identified the various 
regimes, we turn to the evolution of the flow curves with volume fraction. 
The value of the viscosity $\eta_T$ in the thermal Newtonian regime increases 
rapidly with $\phi$ as the glass transition density is approached, 
with $\phi_g \approx 0.59$~\cite{tom}, and the thermal regime is characterized 
by a strong shear-thinning behaviour, $\eta \propto \gdot^{-1}$,
as is typical for a glassy material with a finite yield stress.

The value of the second Newtonian viscosity $\eta_0$ in the athermal 
regime at larger $\gdot$ also displays a strong 
increase with volume fraction, but with a density dependence
distinct from the one of $\eta_T$, 
diverging near the jamming transition at $\phi_J 
\approx 0.64$~\cite{olsson}. 

An interesting behaviour is observed when 
the jamming transition is approached because $\eta_0$ increases 
more rapidly than the value of the viscosity maximum 
observed at the end of the shear-thickening regime. 
At a result, close enough to the 
jamming density these two values become equal, 
which implies that the shear-thickening regime eventually disappears as 
the volume fraction is larger than the jamming density,
see Fig.~\ref{fig1}c.
Indeed, flow curves simplify above jamming and we simply observe 
shear-thinning behaviour for $\phi \gtrsim 0.64$.
In this density regime, the system behaves
as a jammed athermal assembly of soft particles, and is therefore
characterized by a finite yield stress~\cite{ikeda,olsson} which entirely 
dominates the flow curve, such that again $\eta \propto \gdot^{-1}$. 

A comparison to experimental flow curves in Figs.~\ref{fig1}a,b 
shows that all features described in the simulations are also 
present in experiments. The thermal Newtonian and 
shear-thinning regimes, and the glass transition physics are 
observed in the colloidal latex dispersion at small shear rates (small 
P\'eclet number). 
The athermal Newtonian viscosity is observed in both the colloidal
dispersion at large enough $\gdot$ (large P\'eclet number), 
and in the emulsion where the large droplet size ensures that thermal 
fluctuations are irrelevant. 
The strong continuous shear-thickening is also observed 
in both systems at larger shear rates, with a viscosity maximum 
also observed in some cases. 
Finally, when the soft emulsion is compressed above jamming,
shear-thickening is not observed anymore. All these 
features are in excellent agreement with our numerical 
observations, as claimed in the introduction.

\section{Scaling analysis of shear thickening}
   
We now focus on the specific deviations brought about by the inclusion
of inertia into the equation of motion Eq.~(\ref{eom}).
To simplify the discussion, it is useful to analyse the case where
thermal fluctuations are completely absent, $T^\star = 0$, 
such that the competition is between viscous dissipation, 
inertial effects, and particle softness in the vicinity of 
the jamming transition. In other words, we consider the physics
at large P\'eclet number. 

\begin{figure}
\psfig{file=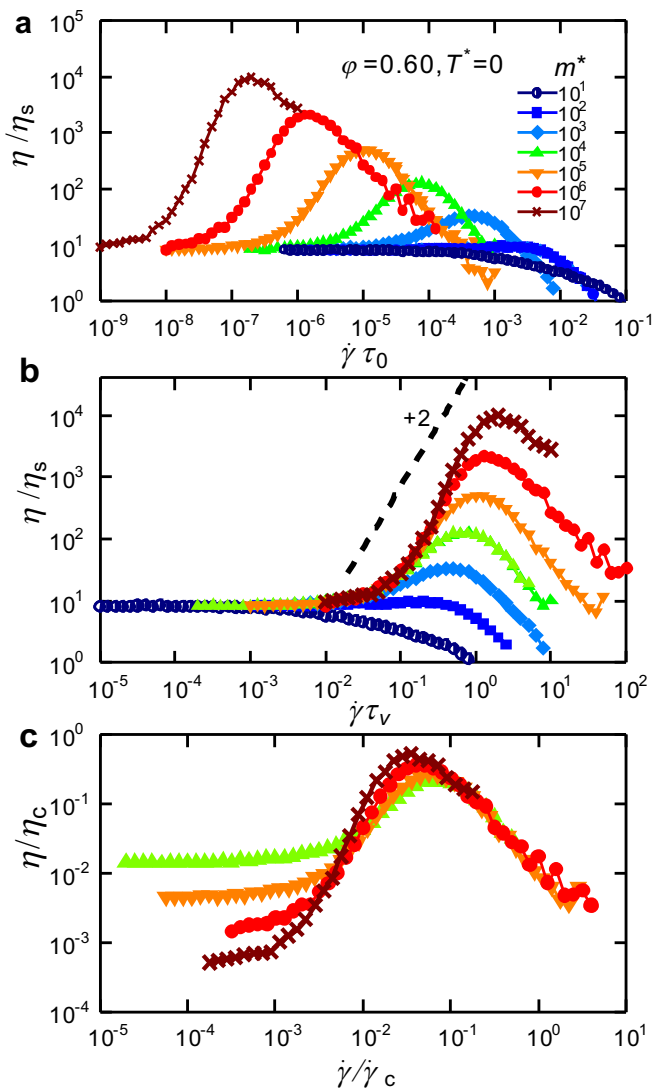,width=8.5cm,clip}
\caption{\label{fig2} 
Flow curves obtained zero temperature, $T^\star = 0$, 
and $\varphi=0.6$ for increasing mass $m^\star$.
(a) Raw viscosity data indicate the emergence of 
shear-thickening with on onset and magnitude controlled 
by inertial terms. 
(b) Same data plotted as a function of the rescaled 
flow rate $\gdot \tau_v$ display a good collapse 
of the onset of shear-thickening near $\gdot \tau_v \approx 10^{-2}$, 
followed by a regime with $\eta \propto \gdot^2$, see Eq.~(\ref{bagnold}).   
(c) Scaling plot of the viscosity maximum 
with parameters taken from Eq.~(\ref{peak}). }
\end{figure}

We first concentrate on a fixed volume fraction 
below the jamming transition, $\phi = 0.6 < \phi_J$,
and study the evolution of the flow curves as 
$m^\star$ is varied over a broad range, 
see Fig.~\ref{fig2}a. For the smaller value, $m^\star = 10$, 
inertia has a negligible effect on the flow curves, which thus resemble
the result obtained in the overdamped limit $m^\star=0$, namely a Newtonian
plateau with viscosity $\eta_0$, followed by shear-thinning at large $\gdot$. 
When $m^\star$ is increased, inertial effects set in and 
the shear-thickening behaviour and viscosity maximum become 
apparent. Clearly, the onset of shear-thickening, the magnitude
of the viscosity increase, and the viscosity maximum are strongly
dependent on $m^\star$, showing that they directly result from inertial effects.

In Fig.~\ref{fig2}b, we show the same data using a rescaled shear rate
$\gdot \tau_v$, using the inertial timescale $\tau_v = m/\xi$ defined above.
Because $\tau_v$ is the typical relaxation time of 
particle velocities due to the viscous damping, 
the condition $\gdot \tau_v \gg 1$ means that shear deformation 
occurs faster than velocity relaxation. In this rescaled plot, 
the onset of shear-thickening in the flow curves at various $m^\star$ collapses 
very nicely, such that $\eta \approx \eta_0$ when $\gdot \tau_v \lesssim 
10^{-2}$, while the data suggest
$\eta \propto \gdot^2$ when  $\gdot \tau_v \gtrsim 
10^{-2}$. While the scaling variable $\gdot \tau_v$ is a
straightforward choice (justified below), the very small value it takes at 
the crossover, $\gdot \tau_v \approx 10^{-2}$, is less intuitive. 

We now rationalize the observed behaviours 
by developing theoretical arguments in the spirit of kinetic theory.
The sole source of dissipation in the Langevin dynamics 
Eq.~(\ref{eom}) is the viscous damping. Energy balance between
dissipation and energy injection by the flow per 
unit volume and unit time yields
\be 
\sigma \gdot = \rho L^3 \times \xi \bar{v}^2 / L^3, 
\label{ec}
\ee
where $\bar{v}$ is the typical amplitude of the particle velocity.
Using a kinetic theory argument, 
the shear stress can also be expressed as the product of 
a number of collisions with a typical momentum transfer at collisions. 
Neglecting the density dependence of the mean free path, this 
gives us a second (approximate) relation:
\be 
\sigma \approx \rho L^2 \bar{v} \times m \gdot a / L^2. 
\label{kt}
\ee
By combining Eqs.~(\ref{ec}, \ref{kt}) we obtain:
\be 
\sigma \approx (\rho m^2 a^2/\xi) \gdot^3, \ \ \ \bar{v} \approx 
(m a / \xi ) \gdot^2. 
\label{bagnold}
\ee
Remarkably, the obtained constitutive equation between stress and shear rate 
fully agrees with the observation $\eta/\eta_s \propto  (\tau_v \gdot)^2$
in Fig.~\ref{fig2}b. Note that this result is 
is different from Bagnold scaling $\eta \propto \gdot$~\cite{bagnold},  
because the shear rheology in this regime strongly depends on the details 
of the energy dissipation~\cite{teitel}. Bagnold scaling is obtained when  
energy dissipation in Eq.~(\ref{fig1}) is introduced 
via collisions, such that the energy balance equation 
in Eq.~(\ref{ec}) is modified into $\sigma \gdot \propto \bar{v}^3$. 
Together with Eq.~(\ref{kt}), Bagnold scaling $\eta \propto 
\bar{v}^2$ would be recovered~\cite{mitaraisan}. 

We can extend our kinetic argument to account for the effect of 
particle softness and explain the origin of the viscosity maximum.
The above analysis suggests that particles move faster 
as $\gdot$ increases, see Eq.~(\ref{bagnold}). 
This implies that `collisions' become ill-defined when
particles have enough kinetic energy to overcome 
their repulsive interactions, which happens when $\bar{v} > \bar{v}_c$ with the 
crossover velocity $\bar{v}_c$ given by: $m \bar{v}_c^2/2 \approx  \epsilon$. 
Combined with Eq.~(\ref{bagnold}) this argument provides a rough 
estimate of the height and location of the viscosity maximum, namely 
\be 
\eta_c = \rho a (m \epsilon)^{1/2}, \ \ \  \gdot_c = [ \epsilon 
\xi^2 / (m^3 a^2) ]^{1/4}.
\label{peak}
\ee 
In Fig.~\ref{fig2}c, the renormalized flow curves
$\eta/\eta_c$ versus $\gdot/\gdot_c$ are shown, where 
the height and location of the viscosity maxima 
are nearly collapsed, suggesting that our crude argument 
captures its origin, namely, the competition between 
inertial effects and particle softness. For collisional
dissipation and soft particles, we predict again a viscosity 
maximum~\cite{preparation} but with different scaling properties controlled 
by $\eta_c = \rho a (m \epsilon)^{1/2}$ and 
$\gdot_c = [\epsilon /(m a^2)]^{1/2}$. 

\section{Dynamic state diagram}
  
\begin{figure}
\psfig{file=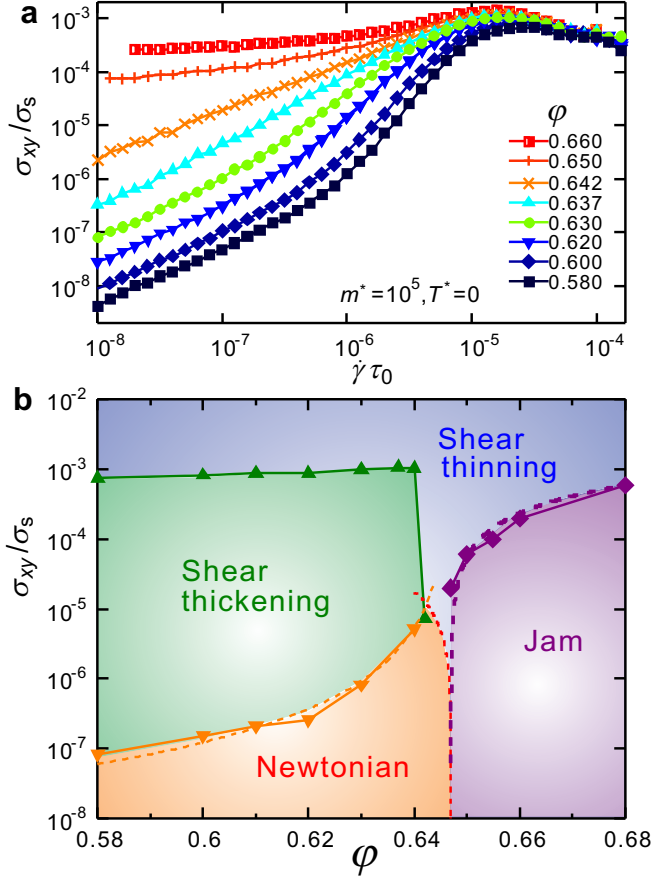,width=8.5cm,clip}
\caption{\label{fig3}
(a) Evolution with volume fraction of the flow curves obtained for 
$T^\star = 0$ and $m^\star =10^5$.
(b) Dynamic state diagram for the same parameters 
showing stress regions where Newtonian, shear-thickening and 
shear-thinning regimes can be observed. The system 
does not flow in the `jam' region, where the stress is below
the yield stress value.}
\end{figure}

So far, we analysed the shear-thickening at constant density. 
To obtain a dynamic state diagram of the multiple rheological 
regimes as determined experimentally~\cite{brownprl}, we turn to the 
influence of the volume fraction. We concentrate again on the 
zero-temperature case, for convenience. In Fig.~\ref{fig3}a, 
we show flow curves with $T^\star=0$, $m^\star =10^5$, and various 
volume fraction from $\phi=0.58$ below jamming up to $\phi=0.66$ above,
while Fig.~\ref{fig3}b summarizes these results in a 
stress / volume fraction dynamic state diagram~\cite{brownprl}. 

The data in Fig.~\ref{fig3}a suggest that  
the onset of shear-thickening is only very weakly dependent
on the volume fraction, and occurs when $\gdot \tau_v \approx 10^{-2}$
independently of $\phi$. (We have validated this hypothesis for a 
much broader range of control parameters~\cite{preparation}.) 
This indicates that the volume fraction dependences 
of the Newtonian viscosity $\eta_0$ and of the inertial regime $\eta \propto
\gdot^2$ are actually identical, suggesting that both regimes 
actually reflect the fundamental flow properties of an athermal assembly 
of hard spherical particles below jamming.
Accordingly, the stress scale controlling 
the Newtonian-to-thickening transition in Fig.~\ref{fig3}b 
is given by $\sigma \approx \eta_0(\phi) \times 10^{-2}/\tau_v$, 
and its evolution with $\phi$ is thus essentially controlled by 
the one of the viscosity $\eta_0(\phi)$, which diverges at $\phi_J$. 
Our simulations also indicate that the density dependence of the 
stress maximum is nonsingular, see Fig.~\ref{fig3}a. 
The stress scale delimiting the upper boundary of the 
shear-thickening region in Fig.~\ref{fig3}b is thus nearly constant. 

When the jamming transition is approached, both Newtonian
and shear-thickening regimes disappear. The former because 
it becomes easier to enter the shear-thinning regime as 
the viscosity $\eta_0$ increases~\cite{olsson}. 
The latter regime disappears
when the viscosity in the Newtonian plateau becomes larger 
than the viscosity maximum at the end of the thickening regime, 
which happens at a density slightly below jamming. 
These two stress boundaries are reported in the diagram of 
Fig.~\ref{fig3}b, which shows that the jammed system at $\phi > \phi_J$
either does not flow when $\sigma$ is below the yield stress (which 
increases continuously with density above $\phi_J$~\cite{ikeda}), 
or displays shear-thinning. The overall structure of 
the state diagram is similar to the experimental results~\cite{brownprl}. 

We have shown that a simple Langevin model of soft repulsive particles 
displays a rheology in good agreement with the complex 
rheology observed in dense suspensions, due to the timescale 
competition between $\tau_0$, $\tau_T$ and $\tau_v$. 
Our simulations indicate in particular that the onset of shear-thickening
occurs when $\gdot \tau_v \approx 10^{-2}$, which agrees excellently
with the value $\gdot \tau_v \sim 0.7 \cdot 10^{-2}$ obtained 
in Fig.~\ref{fig1}b for the emulsion, while a 
somewhat smaller value $\gdot \tau_v \sim 0.4 \cdot 10^{-3}$ 
is found for the latex dispersion shown in Fig.~\ref{fig1}a. We note 
that the addition of frictional forces might affect the nature 
of the shear-thickening onset~\cite{seto,seto2}, but not 
necessarily the shear rate where it occurs~\cite{fernandez}.

\end{document}